\documentclass{pasj00}
\draft
\usepackage{natbib}

\begin{document}
\SetRunningHead{S. Katsuda, et al.}{Suzaku Detection of Diffuse Hard
  X-Ray Emission outside Vela X} 
\Received{2011/2/4}
\Accepted{2011/3/24}

\title{Suzaku Detection of Diffuse Hard X-Ray Emission outside Vela X} 

%
 \author{%
   Satoru \textsc{Katsuda}\altaffilmark{1}, Koji
   \textsc{Mori}\altaffilmark{2}, Robert
   \textsc{Petre}\altaffilmark{1}, Hiroya 
   \textsc{Yamaguchi}\altaffilmark{3}, Hiroshi
   \textsc{Tsunemi}\altaffilmark{4}, Fabrizio
   \textsc{Bocchino}\altaffilmark{5}, Aya 
   \textsc{Bamba}\altaffilmark{6,7}, Marco 
   \textsc{Miceli}\altaffilmark{8,5}, John W. 
   \textsc{Hewitt}\altaffilmark{1}, Tea
   \textsc{Temim}\altaffilmark{1}, Hiroyuki
   \textsc{Uchida}\altaffilmark{4}, and Rie
   \textsc{Yoshii}\altaffilmark{3}
}
 \email{Satoru.Katsuda@nasa.gov}
 \altaffiltext{1}{Code 662, NASA Goddard Space Flight Center,
 Greenbelt, MD 20771, U.S.A.} 
\altaffiltext{2}{Department of Applied Physics, Faculty of Engineering,
University of Miyazaki, 889-2192, Japan}
\altaffiltext{3}{RIKEN (The Institute of Physical and Chemical
  Research), 2-1 Hirosawa, Wako, Saitama 351-0198}
\altaffiltext{4}{Department of Earth and Space Science, Graduate
   School of Science, Osaka University, 1-1 Machikaneyama, Toyonaka,
   Osaka 560-0043, Japan}
\altaffiltext{5}{INAF -- Osservatorio Astronomico di Palermo, Piazza
   del Parlamento 1, 90134 Palermo, Italy} 
\altaffiltext{6}{School of Cosmic Physics, Dublin Institute for
   Advanced Studies, 31 Fitzwilliam Place, Dublin 2, Republic of
   Ireland} 
\altaffiltext{7}{ISAS/JAXA Department of High Energy Astrophysics,
   3-1-1 Yoshinodai, Chuo-ku, Sagamihara, Kanagawa 252-5210, Japan} 
\altaffiltext{8}{Dipartimento di Scienze Fisiche ed Astronomiche,
   Sezione di Astronomia, Università di Palermo, Piazza del Parlamento
   1, 90134 Palermo, Italy} 


\maketitle

\begin{abstract}
Vela X is a large, 3$^{\circ}\times2^{\circ}$, radio-emitting 
pulsar wind nebula (PWN) powered by the Vela pulsar in the Vela 
supernova remnant.  Using four Suzaku/XIS observations pointed
just outside Vela X, we find hard X-ray emission extending throughout
the fields of view.  The hard X-ray spectra are well represented by 
a power-law.  The photon index is measured to be constant
at $\Gamma\sim2.4$, similar to that of the southern outer part of
Vela X.  The power-law flux decreases with increasing distance from
the pulsar.  These properties lead us to propose that the hard X-ray 
emission is associated with the Vela PWN.  The larger X-ray extension
found in this work strongly suggests that distinct populations
relativistic electrons form the X-ray PWN and Vela X, as was recently
inferred from multiwavelength spectral modeling of Vela X.  
\end{abstract}

\section{Introduction}

It has been universally accepted that a considerable fraction of
a pulsar's spin-down energy is converted into an outflow of
relativistic particles.  This outflow is terminated by a strong shock
resulting when the relativistic particles interact with ejecta from
the host supernova remnant (SNR).  High-energy particles further
accelerated at this shock and diffusing downstream emit radio to TeV
$\gamma$-rays via synchrotron or inverse Compton processes.  The
non-thermal emission is observed as a pulsar wind nebula (PWN). 

One key event occurring during the evolution of a PWN (Gaensler \&
Slane 2006 and references therein) is its interaction with the reverse
shock propagating back into the SNR interior.  It is assumed that this
interaction causes magnetic-field amplification and hence rapid
synchrotron cooling of the highest energy (TeV) relativistic
electrons.  Since the cooling time of GeV electrons responsible for
the radio emission is considerably longer (the lifetime of
relativistic electrons inversely depends on their energy), a radio
relic PWN is left behind for a long time after this interaction. 

\begin{figure}[h]
  \begin{center}
    \FigureFile(150mm,200mm){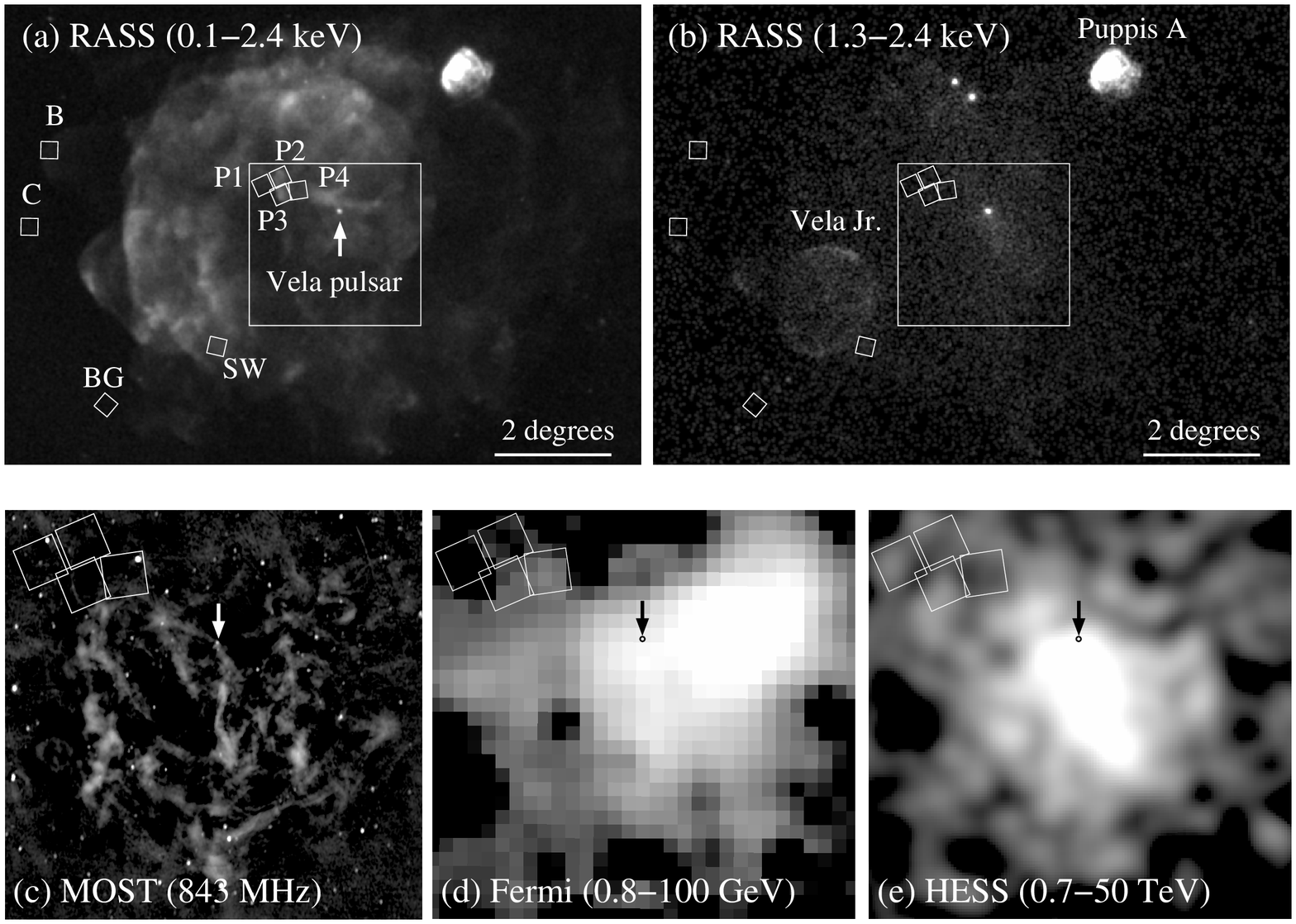}
  \end{center}
  \caption{(a) RASS X-ray (0.1--2.4\,keV) image of the entire Vela
    SNR. The effects of vignetting and exposure are corrected.  The
    intensity scale is square root.  The Suzaku/XIS FOV
    (\timeform{17'.8}$\times$\timeform{17'.8}) are shown as white
    boxes labeled P1--P4 (for source regions), BG (for local
    background), B, C, and SW (for Vela shrapnel B and C, and outside 
    the SW edge of Vela Jr., respectively).
    (b) Same as (a) but for the 1.3--2.4\,keV band.  Galactic SNRs
    Puppis~A and Vela Jr.\ superposed along the line of sight can be
    seen in the NE edge or SE corner of the Vela SNR. 
    (c) MOST radio (843\,MHz) image covering Vela X region (the white
    boxes in figure~\ref{fig:all_image} (a) and (b)), with its network
    of filaments (Bock et al.\ 1998).  The intensity scale is square
    root.  The XIS FOV are located just outside Vela X.  The arrow
    indicates the location of the Vela pulsar. 
    (d) Same as (c) but for Fermi (0.8--100\,GeV) image, which a 
    test statistic map as in Abdo et al.\ (2010).
    (e) Same as (c) but for HESS (0.7--50\,TeV) image, which is taken 
    from Aharonian et al.\ (2006).
  } 
    \label{fig:all_image} 
\end{figure}

Of the more than 50 known PWNe (e.g., Kargaltsev \& Pavlov 2008),
the Vela PWN, associated with the middle-aged (11000\,yr: Reichley et
al.\ 1970) Vela SNR (see, figure~\ref{fig:all_image}), is located the 
nearest to Earth ($\sim$290\,pc: Dodson et al.\ 2003).  This proximity
makes it an ideal target for detailed spatial studies.  With its
superior (sub-arcsecond) spatial resolution, Chandra X-ray Observatory
revealed spectacular small-scale ($\sim$\timeform{10''}) structures
within the Vela PWN, including jets and arcs around the pulsar
(Helfand et al.\ 2001;  Pavlov et al.\ 2001).  These features are
surrounded by a bright compact nebula emitting non-thermal X-rays, a
``kidney-bean'' nebula of size $\sim$2$^{\prime}$ (Harnden et al.\
1985).  The spectra of these features are characterized by a power-law
component, with photon index varying from $\Gamma\sim1$ for the inner
jets and arcs to $\Gamma\sim1.5$ for the surroundings (e.g.,
Kargaltsev \& Pavlov 2008).  This system is further embedded in a
large region emitting non-thermal X-rays whose power-law index
($\Gamma\sim2$) is steeper than those in the inner regions (Markwardt
\& \"Ogelman 1995; 1998; Mangano et al.\ 2005; LaMassa et al.\ 2008).
However, the full morphology of this non-thermal X-ray--emitting
region (X-ray PWN) is not yet known, because Vela has not been mapped
by a sufficiently broad band imaging spectrometer.

It is important for the study of the evolution of the Vela PWN and
PWNe in general to map and understand the large-scale
($\sim1^{\circ}$) structure of the Vela PWN across the electromagnetic
spectrum.  In the radio band, the Vela PWN is a network of bright
filaments encompassing a large, 3$^\circ\times$2$^{\circ}$, region
around the pulsar as shown in figure~\ref{fig:all_image} (c).  It is
referred to as Vela X (Frail et al.\ 1997; Bock et al.\ 1998 and
references therein).  Vela X is asymmetric relative to the pulsar.
Both the chaotic filamentation and the asymmetry led Blondin et al.\
(2001) to propose that Vela X arose from interaction between the PWN
and the reverse shock of the Vela SNR in an inhomogeneous ambient
medium.  Thus, the Vela PWN is regarded as a prototype radio relic PWN
(Gaensler \& Slane 2006).  Recently, observations using HESS, AGILE,
and Fermi have revealed TeV (figure~\ref{fig:all_image} (d)) and GeV
$\gamma$-ray (figure~\ref{fig:all_image} (e)) maps (Aharonian et al.\
2006; Pellizzoni et al.\ 2010; Abdo et al.\ 2010).  The TeV/GeV
morphologies are also asymmetric relative to the pulsar, whereas they
are quite different with each other, showing chaotic structures of the
Vela PWN. X-ray mapping of the Vela PWN is now ongoing.  An important
point is that the lifetime of relativistic electrons emitting X-rays
is the shortest, because they are the most energetic (assuming the GeV
and TeV emission is inverse Compton radiation).  Therefore, the X-ray 
morphology might be different from Vela X, if Vela X is really a
long-lived radio relic.

We here report on the detection of X-ray synchrotron emission beyond
the boundary of Vela X, based on Suzaku/XIS data with unprecedented
high sensitivity in the hard X-ray energy band (2--10\,keV).  Our
analysis indicates that this emission is associated with the Vela PWN,
strongly suggesting that the morphology of Vela PWN in X-rays is  
different from that in the radio (i.e., Vela X).  We suggest that
the X-ray PWN and Vela X are formed by distinct populations
relativistic electrons, consistent with the picture inferred from
recent multiwavelength spectral modeling of Vela X (LaMassa et al.\
2008; de Jager et al.\ 2008).  This implies that Vela X is indeed a
radio relic of the interaction between the Vela PWN and the reverse
shock.

\section{Observations and Data Screening}

We observed part of the Vela SNR with {\it Suzaku} (Mitsuda et al.\ 
2007) in four pointings during 2010 May 2--3.  We here
concentrate on the data taken by the X-ray Imaging Spectrometer 
(XIS: Koyama et al.\ 2007).  The XIS consists of two
front-illuminated (FI: XIS0 and XIS3) CCD cameras and one
back-illuminated (BI: XIS1) CCD camera.  Each camera covers an
identical imaging area of \timeform{17'.8}$\times$\timeform{17'.8}. 
Figure~\ref{fig:all_image} (a) shows the four XIS fields of view
(FOV) as white boxes labeled from P1 to P4 on a 
Vela SNR image from the ROSAT All-Sky Survey (RASS).  Another box
labeled BG shown in the figure represents the XIS FOV where we
estimate local X-ray background for our spectral analysis.  
Figure~\ref{fig:all_image} (b) shows a hard-band RASS image covering
the same area in figure~\ref{fig:all_image} (a), from 
which the Galactic SNRs Puppis~A and Vela Jr.\ superposed along the
line of sight appear as bright extended regions at the northwestern
(NW) and southeastern (SE) corners of the remnant, respectively,
whereas the Vela PWN is not apparent. 
Figure~\ref{fig:all_image} (c) presents a radio (843 MHz) image of 
the central region of the Vela SNR.  The network of filaments which
forms Vela X is clearly seen in the figure.  Our XIS FOV are located
just outside Vela X to the northeast (NE).

Using the latest CALDB files (e.g., a CTI calibration file of version 
20091202), we reprocessed and cleaned the data according to the standard 
criteria\footnote{See the Suzaku Data Reduction Manual which can be
  found from http://heasarc.gsfc.nasa.gov/docs/suzaku/analysis/abc.}
recommended by the calibration team of Suzaku/XIS.  After the 
screening, the remaining exposure times are 25.3\,ks, 23.2\,ks, 
16.8\,ks, and 16.8\,ks for P1, P2, P3, and P4, respectively.

\section{Analysis and Results}

\begin{figure}[h]
  \begin{center}
    \FigureFile(150mm,150mm){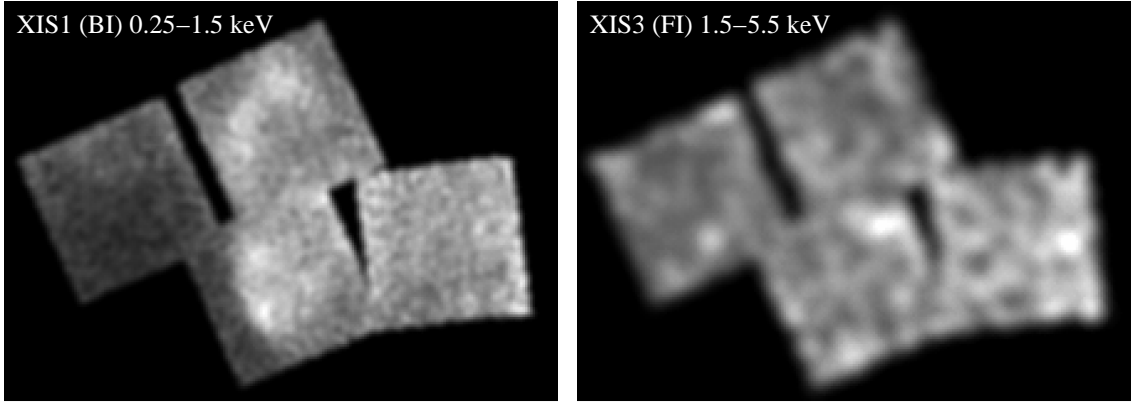}
  \end{center}
  \caption{Left: Linearly-scaled XIS1 image in an energy range of
        0.25--1.5\,keV.  The effects of vignetting and exposure are
	corrected after subtraction of the NXB emission.  The image
        has been smoothed with a Gaussian kernel of $\sigma$=50$''$.
	Two corners of each FOV.  Right: Same as left but
	obtained by XIS3 in the 1.5--5.5\,keV band, and 
	smoothed with a Gaussian kernel of $\sigma$=117$''$.
        }
	\label{fig:xis_image} 
\end{figure}

Mosaics of the XIS images are shown in figure~\ref{fig:xis_image}.
The left and right panels show the soft (0.25--1.5\,keV) and hard
(1.5--5.5\,keV) X-ray band images, respectively, with vignetting-corrections
subtraction of the non X-ray background (NXB) caused by charged
particles and $\gamma$-rays hitting the detectors (Tawa et al.\
2008). Two corners of each FOV, where the calibration source of
$^{55}$Fe is illuminated, are masked in these images.  We see strong
contrasts in the soft band image (left panel).  As we discuss below,
this contrast is most likely caused by inhomogeneities of thermal 
emission from the Vela SNR.  Figure~\ref{fig:xis_image} right shows hard
X-ray emission extending throughout the FOV.  This emission is at
least a factor of 2 above the local background level. 

\begin{figure}[h]
  \begin{center}
    \FigureFile(150mm,150mm){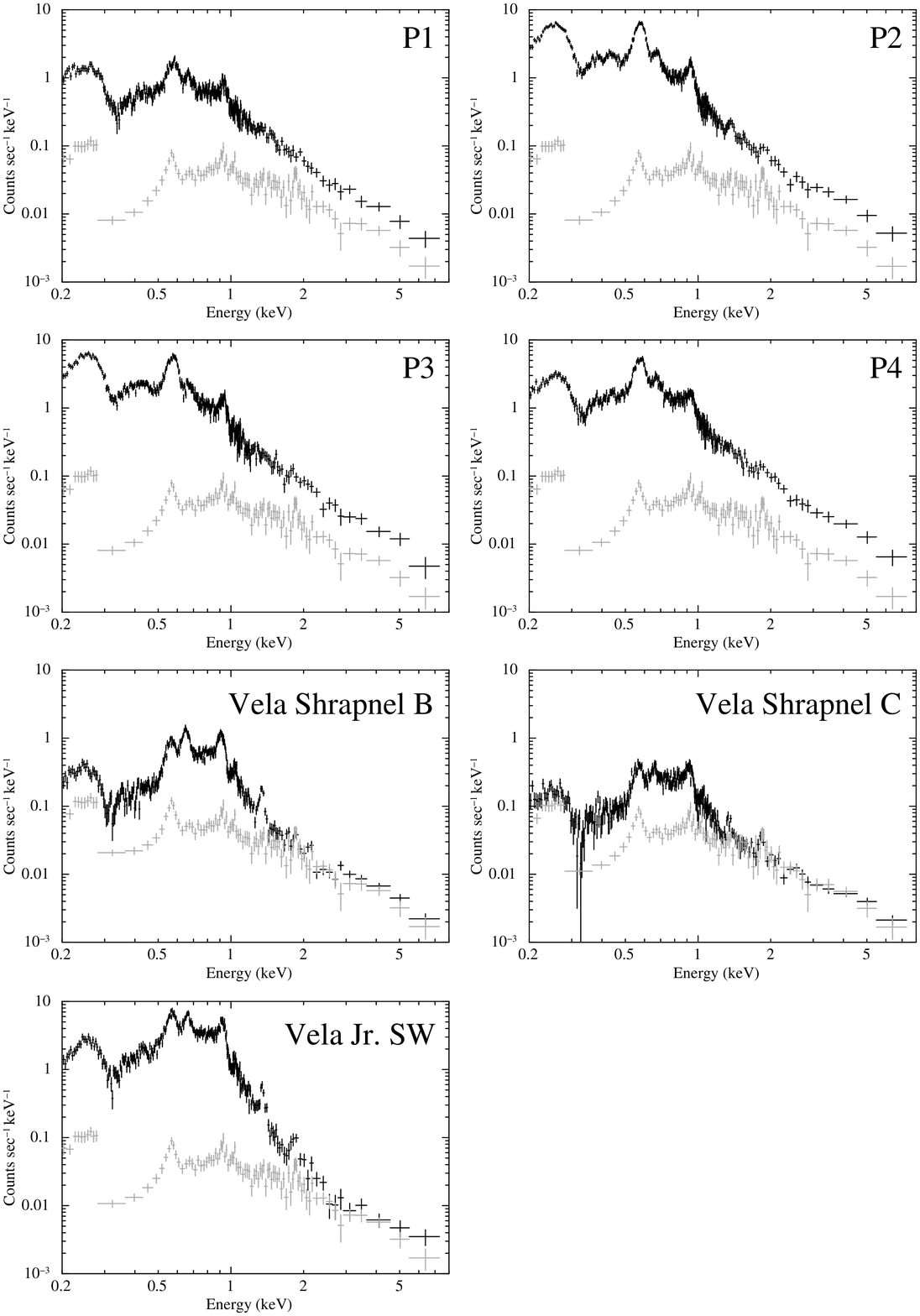}
  \end{center}
  \caption{NXB-subtracted source spectra (in black) and a local 
	background (in gray).  The spectra are extracted from the
	entire FOV.  The background spectra are corrected for the
	degradation of quantum efficiency after NXB subtraction. 
	}
	\label{fig:spec_withbg} 
\end{figure}

As the first step of our spectral analysis, we subtract the NXB 
(generated by the {\tt nxbgen} software: Tawa et al.\ 2008) from the
source spectra.  Figure~\ref{fig:spec_withbg} shows the NXB-subtracted 
XIS1 spectra (in black) extracted from the entire FOV for each 
observation.  Then, the X-ray background is obtained from the local 
background region indicated in figure~\ref{fig:all_image}.  The
NXB-subtracted local-background XIS1 spectra are plotted as gray
crosses in figure~\ref{fig:spec_withbg}.  For these, we take into
account the degradation of quantum efficiency due to the build-up of
contaminants on the optical blocking filter of the XIS.  A detailed
description of this correction can be found in the literature
(e.g., Yamaguchi \& Katsuda 2009).  Figure~\ref{fig:spec_withbg}
reveals the presence of featureless continuum emission above the
background level in the hard energy band ($>$1.5\,keV), whereas the
spectra below $\sim$1.5\,keV are dominated by line emission such as O
He$\alpha$ at $\sim$0.57\,keV, O Ly$\alpha$ at $\sim$0.65\,keV, and Ne
He$\alpha$ at $\sim$0.91\,keV.  In figure~\ref{fig:spec_withbg}
bottom, we plot Suzaku spectra of Vela shrapnel B and C, and a region
outside the southwestern (SW) edge of Vela Jr.\ (see,
figure~\ref{fig:all_image}) along with the local background corrected
for effective area.   We see that these spectra do not show hard X-ray
emission, indicating that (1) the hard X-ray emission in our FOV
(P1--P4) does not extend to these regions and (2) the local cosmic
X-ray background is quite uniform around the Vela SNR.

In the following fitting procedure, we use photons in the energy range 
of 0.4--11.0\,keV, and employ the XSPEC software (version 12.6.0q).  
Since the two FI chips (XIS0 and XIS3) have similar spectral
responses, their spectra are summed to improve the photon statistics.
We fit the FI and BI spectra simultaneously.  We manually adjust the
energy scale to obtain better fits, by allowing energy-scale offsets
to vary freely for the FI and the BI detectors, respectively.  Before
fitting, each spectrum is grouped into bins with at least $\sim$20
counts after subtracting the local background, which allows us to
perform a $\chi^2$ test.

\begin{figure}[h]
  \begin{center}
    \FigureFile(150mm,150mm){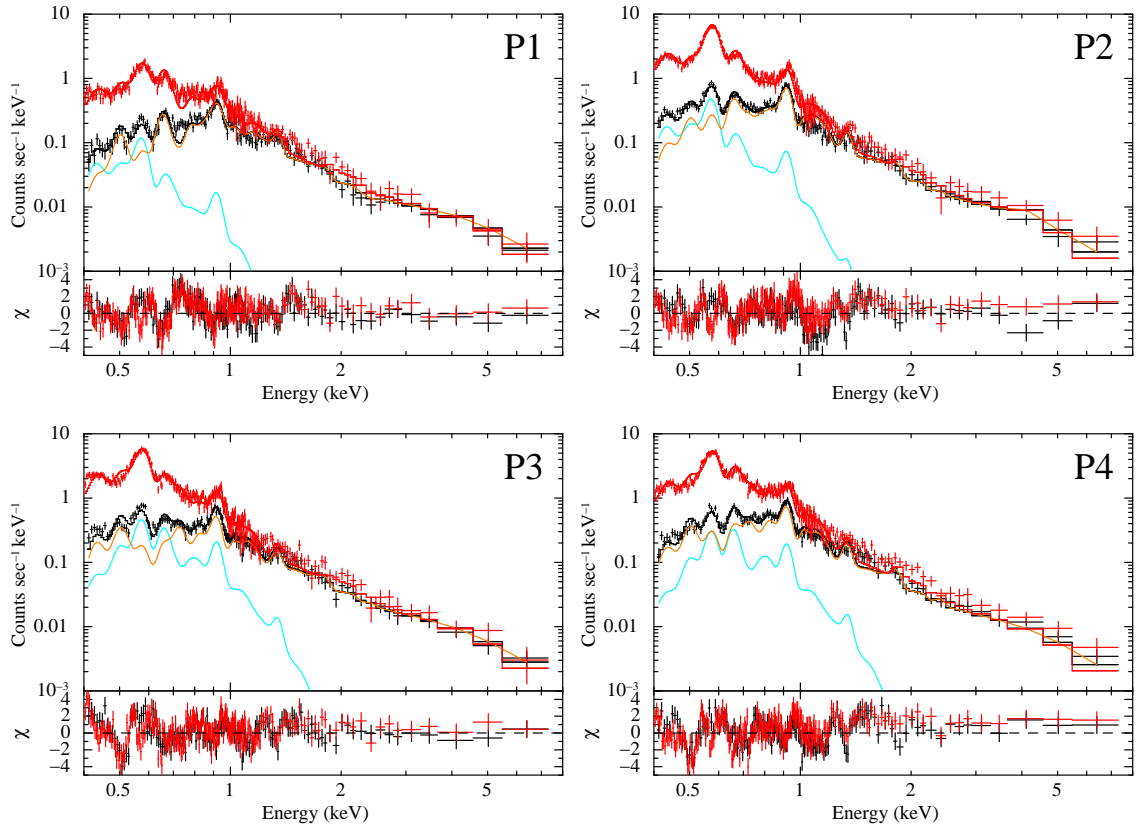}
  \end{center}
  \caption{XIS spectra along with the best-fit model consisting of two
	thermal components.  Red and black correspond to XIS1 and
	XIS0+3, respectively.  Contributions of each component ({\tt
	apec} in light blue and {\tt vnei} in orange) are separately
	illustrated only for XIS0+3.  Lower panels show residuals
	between the data and the best-fit model.
	}
	\label{fig:fit1}
\end{figure}

We follow previous ROSAT studies, using a two-temperature collisional
ionization equilibrium (CIE) plasma model (Bocchino et al.\ 1997;
1999; Lu \& Aschenbach 2000; Miceli et al.\ 2005). It should be noted,
however, that the ROSAT data have inadequate spectral resolution to
reveal non-equilibrium ionization (NEI), but that NEI conditions
cannot be rejected by the data (Bocchino et al.\ 1999; Miceli et al.\
2005). In fact, recent X-ray CCD observations of several locations in
the Vela SNR have revealed NEI conditions  (e.g., Tsunemi et al.\
1999; Katsuda \& Tsunemi 2005; Yamaguchi \& Katsuda 2009).  We
therefore employ a two-component thermal plasma model that combines
NEI and CIE conditions: the hotter component (hereafter, T1) uses an
NEI model, whereas the cooler component (hereafter, T2) is assumed to
be in CIE, because the data cannot distinguish for it between NEI and
CIE.  We adopt the {\tt tbabs} model (Wilms et al.\ 2000) for the
absorption, either {\tt vnei} or {\tt  vpshock} model (Borkowski et 
al.\ 2001) with variable abundances for T1, and the {\tt apec} model
(Smith et al.\ 2001) with solar abundances for T2, respectively.
The {\tt vnei} model has a single ionization timescale ($\tau =
n_\mathrm{e}t$, where $n_\mathrm{e}$ is the electron density and $t$
is the time after the shock heating), whereas the {\tt vpshock} model
assumes a range of $\tau$, which we take 0 to the fitted maximum
value.  In these thermal emission models, the solar abundances are
based on Anders \& Grevesse (1989).  When applying this model for each
spectrum from the four fields (P1--P4), we obtain the best-fit models
shown in Figure~\ref{fig:fit1}. We find that the models are rejected
with high confidence (reduced $\chi^2$ of $\sim$2).  We also find
unusually high electron temperatures of a few keV and very low metal
abundances below 0.1 solar values for the high-temperature component. 

Addition of a power-law component resolves the uncomfortable physical 
conditions and results in acceptable fits (formally speaking,
acceptable only if we consider systematic uncertainties).  
The addition of a thermal component with an electron temperature of a 
few keV, instead of the power-law component, also yields satisfactory
fits.  However, again, such a high-temperature component is not
expected for evolved SNRs like Vela (11000\,yr).  Furthermore, the 
thermal component shows implausible metal abundances of $\sim$0.01
solar values.  Therefore, the thermal interpretation is unlikely to be
correct, and we therefore accept the non-thermal power-law
interpretation.  Table~\ref{tab:fit_param} summarizes the fit details.
Figure~\ref{fig:fit1} shows the XIS spectra along with the best-fit
models of {\tt tbabs}$\times$({\tt vnei} + {\tt apec} +
{\tt power-law}) and {\tt tbabs}$\times$({\tt vpshock} + {\tt apec} +
{\tt power-law}).  We find that the results from {\tt vnei} and {\tt 
vpshock} (for T1) are consistent with each other.

Possible sources for the power-law component would be either the
extended Vela PWN, relativistic particles accelerated in the Vela
SNR's shell, or unrelated background possibly connected to Vela Jr.
The large distance between Vela Jr.\ and our XIS FOV (see,
figure~\ref{fig:all_image} (b)) make it difficult for emission from
Vela Jr.\ to be present in our FOV.  In addition, as can be found in
table~\ref{tab:fit_param}, the relatively flat photon index of
$\Gamma\sim2.4$ may rule out an association with the shell of Vela
Jr.\ which shows a steeper photon index of $\Gamma\sim2.8$ (e.g.,
Hiraga et al.\ 2009), though a population of relativistic particles
escaping from Vela Jr.\ would have a flatter spectrum as lower-energy
particles would be unable to diffuse as far ahead.  It is also
unlikely that the evolved Vela SNR show such a flat photon index, if
non-thermal emission is present.  Although we cannot fully exclude the
possibility that the hard X-ray emission is another PWN along the line
of sight, the most plausible interpretation seems to be an association
with the Vela PWN.  We will further discuss this interpretation in the
next section.

The power-law flux gradually increases from P1 to P4 by a factor of
$\sim$2.  The photon index is almost constant at $\Gamma\sim$2.4.  One
may worry about possible coupling between the photon index and other
parameters.  To this end, we investigate confidence contours of the
photon index against the electron temperature in T1 (using {\tt
  vnei}), the emission measure in T1 (using {\tt vnei}), and the
normalization in the power-law component.  Figure~\ref{fig:conf_cont}
shows the confidence contours.  We find conservative 90\% confidence
ranges of the photon index in the four FOV to be 2.2--2.6.  Metal
abundances are generally sub-solar and their relative abundances are
consistent with the solar ratios within a factor of 2.  The inferred
electron temperatures ($\sim$0.1\,keV and $\sim$0.3\,keV) are similar
to those found in the pure ISM components in the northern shell
regions (Miceli et al.\ 2005; 2008).  These results suggest that the
thermal emission in our FOV is not dominated by SN ejecta but by
shock-heated interstellar medium.  The emission measure of T2 varies
significantly from field to field, suggesting interstellar medium
inhomogeneities.  It should be, however, noted that absolute (relative
to H) abundances are difficult to measure, especially when thermal
X-ray continuum is contaminated by nonthermal continuum emission. 
In this context, the ejecta origin for the thermal components cannot
be fully excluded.  In fact, some other regions in the remnant show 
evidence of ejecta; e.g., the closest ejecta-dominated region to our
FOV is the so-called ``cocoon'' located south of the pulsar (LaMassa
et al.\ 2008; Mori et al.\ in preparation).  Revealing the ejecta
distribution in this remnant is a topic for future work.

We here examine the {\tt srcut} model (Reynolds 1998) instead of the
{\tt power-law} model for P4.  In this fitting, we fix the radio
photon index to $\alpha$=0.4, which is a typical value for Vela X
(Alvarez et al.\ 2001).  The radio intensity at 1 GHz is treated as
either a fixed parameter or a free parameter.  For the former (fixed)
case, we use the intensity of 10 Jy, which is inferred from the MOST
image in figure~\ref{fig:all_image} (whose image resolution is
\timeform{43''}: Bock et al.\ 1998) smoothed with Suzaku's spatial 
resolution of \timeform{2'} (Serlemitsos et al.\ 2007).  We allow the
roll-off frequency to vary freely.  The other parameters are treated
as in table~\ref{tab:fit_param}.  We find that the {\tt srcut} model
does not fit as well as the {\tt power-law} model ($\chi^2$ values are
879 or 519 for fixed or free radio intensity, respectively).
Nonetheless, we note that the best-fit radio intensity inferred from
the {\tt srcut} model of the latter case is $\sim$0.01 
Jy, which is three orders of magnitude smaller than that of Vela X.
This might imply that X-ray emission is not associated with Vela X.  

\section{Discussion} 

Utilizing the high sensitivity of the XIS, we found diffuse, hard
X-ray emission in four FOV located $\sim$\timeform{0D.6}--\timeform{1D.6} 
northeast of the Vela pulsar.  They are thus located
just outside Vela X, a network of radio filaments that has been
suggested to be a relic of interaction between the PWN and the
reverse shock.  The hard X-ray spectra are featureless and are well
represented by a power-law component.  Figure~\ref{fig:index_flux}
shows photon index (circles) and flux (triangles) of the power-law
component as a function of the distance from the Vela pulsar, which 
is derived using the {\tt vnei} model for the T1 component (see,
table~\ref{tab:fit_param}).  We see that the photon index is almost 
constant at $\Gamma\sim$2.4, consistent with that found in the
southern outer region of Vela X (LaMassa et al.\ 2008; Mori et al.\ in
preparation).  On the other hand, the power-law flux gradually
decreases with increasing distance from the pulsar.  Moreover, the
power-law component becomes negligible 
at larger distances from the pulsar, including regions in the northern
shell (Miceli et al.\ 2005; 2006; 2008) and the Vela shrapnel to the
east (Tsunemi et al.\ 1999; Miyata et al.\ 2001; Katsuda \& Tsunemi
2005; 2006; Yamaguchi \& Katsuda 2009).  These facts suggest that the
hard X-ray emission found in our FOV is related to the Vela PWN, and
thus the PWN in  X-ray extends beyond the radio PWN (i.e., Vela X).
We note that other evolved PWNe associated with PSR~J1826--1334
(Uchiyama et al.\ 2009) and PSR~J1809--193 (Anada et al.\ 2010) share
the large-scale ($>$ several pc) photon index uniformity, in contrast
to young PWNe where the spectrum hardens toward the pulsars (e.g.,
Mori et al.\ 2004 and references therein).

Our finding that the X-ray Vela PWN extends farther than Vela X
provides us with important insight into the origin of relativistic 
electrons.  The mechanisms that potentially limit PWN size are
synchrotron cooling, diffusion, or convection (advection) of the 
relativistic electrons.  If cooling is important and relativistic
electrons systematically flow outward, the X-ray size must
be smaller than the radio size, since the lifetime of X-ray--emitting  
high-energy (TeV) electrons is much shorter than that of
radio-emitting low-energy (GeV) electrons.  An example for such PWNe
would be the Crab nebula.  In the other cases, the size would be
independent of wavelength; in other words, the size in X-ray and radio
would be identical like 3C~58.  Our result, X-ray--emitting
particles beyond the radio-emitting particles, is inconsistent with 
these expectations.  This difficulty can be resolved, however, if we 
postulate that the X-ray--emitting electrons are not physically
related to those forming Vela X.  In fact, recent multiwavelength
spectral modeling of the non-thermal emission in the cocoon have
suggested an excess of low-energy (GeV) electrons (LaMassa et al.\
2008; de Jager et al.\ 2008; Pellizzoni et al.\ 2010; Abdo et al.\
2010).  Thus, our result provides strong support for the idea that
distinct populations relativistic electrons generate the X-ray PWN and 
Vela X from an independent, morphological point of view.

We here speculate how the X-ray PWN becomes more extended than the 
radio PWN (Vela X).  Numerical simulations show that the 
reverse shock formed in the SNR starts crushing the PWN on a time
scale of $\sim$3000 yr after the SN explosion (Blondin et al.\ 2001;
Bucciantini et al.\ 2003).  During the next $\sim$3000 yr or so,
hydrodynamic instabilities at the contact discontinuity between the SN
ejecta and the PWN create filamentary features, which probably have
amplified magnetic fields, which in turn radiate synchrotron emission
efficiently.  This is the process considered to be the origin of Vela
X.  Given an SNR age of 11000 yr, the age of Vela X would therefore be
roughly 5000 yr.  Because the lifetime of radio-emitting electrons is
longer than the age of the Vela SNR, it is possible that these
filamentary features have persisted as Vela X since its birth (Blondin
et al.\ 2001).  On the other hand, the lifetime of the relativistic
electrons emitting synchrotron X-rays is much shorter.  Using the most
recent estimate of the magnetic field B=4\,$\mu$G in the cocoon (and
halo) of Vela X (Abdo et al.\ 2010), the synchrotron cooling time for
electrons emitting synchrotron X-rays with a characteristic energy 
$E_\mathrm{syn}$=2\,keV is $\sim$3400 (B/4
$\mu$G)$^{-1.5}$($E_\mathrm{syn}$/2 keV)$^{-0.5}$ yr (de Jager \& 
Djannati-Ata\"i 2008).  This lifetime is less than the estimated age
of Vela X.  Further, if we consider that magnetic fields were
amplified at the time of interaction between the reverse shock and the 
PWN, then the lifetime is even shorter.  Therefore, most electrons,
which had emitted synchrotron X-rays when Vela X was created, have
already burned-out. In this case, it is reasonable to consider that 
the source of X-ray emission in the Vela PWN is a fresh population of 
high-energy electrons that overran the radio relic (Vela X).  Whereas
this interpretation seems likely, we should keep in mind that the
estimates of the age of Vela X ($\sim$5000\,yr) and the lifetime 
of X-ray--emitting relativistic electrons ($\sim$3400\,yr) are quite
uncertain; both estimates are not as accurate as $\sim$50\%.
Therefore, at this point we cannot rule out other more exotic
possibilities, e.g., somewhat lower-energy electrons at the edge of  
Vela X are now encountering higher magnetic fields for some reason, in
which they can radiate X-rays.

Based on systematic studies of PWNe using Suzaku and Chandra, Bamba et 
al.\ (2010) report that X-ray PWNe continue to expand until an age of 
$\sim$100\,kyr.  Figure 3 in Bamba et al.\ (2010), showing X-ray size
vs.\ characteristic age for 8 PWNe, suggests that the X-ray size 
of the Vela PWN at an age of 11000\,yr should be $\sim$5--10\,pc.  The
farthest edge of our XIS FOV is $\sim$8\,pc away from the pulsar (at a
distance of 290\,pc), within the size range inferred for other PWNe.
To investigate how the X-ray PWN expands to such an extent is left for
future work.  To do this, we need more information on the extent and
morphology of the hard X-rays. 

So far, the only large-scale ($\sim1^{\circ}$) map of the X-ray
synchrotron nebula around the Vela pulsar is the significance-contour
map in 2.5--10\,keV range by Willmore et al.\ (1992) who used the
Birmingham coded mask telescope.  The authors found a diffuse, hard
X-ray feature extending approximately 1$^{\circ}$ to the NE and
SW, and interpreted it to originate from the
Vela PWN.  Its NE edge corresponds to our innermost (the nearest to
the pulsar) XIS FOV, P4, where we find the strongest intensity of hard
X-ray emission among the four (P1--P4) FOV.  Therefore, it appears
that the Suzaku/XIS has unveiled that the PWN extends beyond the NE edge
defined by Willmore et al.\ (1992).  However, we suspect that the
significance-contour map does not reflect a true hard X-ray map.  For
example, the map does not show Vela Jr.\  whose NW rim is
more than 10 times brighter than P4 and thus should have been detected 
(1--5\,keV: Pannuti et al.\ 2010).  Moreover, it should be
noted that the significance level of the suggested PWN feature is not
very high, only 1.5--3 sigma.  In fact, there are a number of such
low-significance hot spots in the map.  Most of these features do not
have counterparts in other wavelengths.  While the authors
interpreted that they are blobs of material at a wide range of
temperatures, an alternative interpretation would be that they are
just statistical noise.  Given these considerations, we believe that
the map presented in Willmore et al.\ (1992) is not
conclusive. Dedicated observations of Vela X and its surroundings with
X-ray CCD spectrometers such as Suzaku/XIS or XMM-Newton/EPIC are
eagerly desired to reveal definitive structures of the X-ray Vela PWN
as well as to investigate the evolution of PWNe in general.

\section{Conclusion} 

We have detected hard X-ray emission just outside Vela X, using four
Suzaku/XIS observations.  Our analysis suggests that it is of
power-law origin and is associated with Vela PWN.  However, the fact 
that the X-ray extent is larger than Vela X strongly suggests that the
hard X-ray emission is not related to Vela X, invoking distinct
populations relativistic electrons for the X-ray PWN and Vela
X---an interpretation recently inferred from multiwavelength spectral 
modelings of Vela X (LaMassa et al.\ 2008; de Jager et al.\ 2008).
Our finding of the large extent of the X-ray PWN would require further
observations to unveil structures of the Vela PWN.

\bigskip

We would like to express our special thanks to Una Hwang for a number
of useful comments, and Douglas Bock and the HESS collaboration for
providing a radio (843 MHz) image and a HESS image of Vela X,
respectively.  S.K.\ is supported by a JSPS Research Fellowship for
Research Abroad, and in part by the NASA grant under the contract
NNG06EO90A.


\begin{figure}[h]
  \begin{center}
    \FigureFile(150mm,150mm){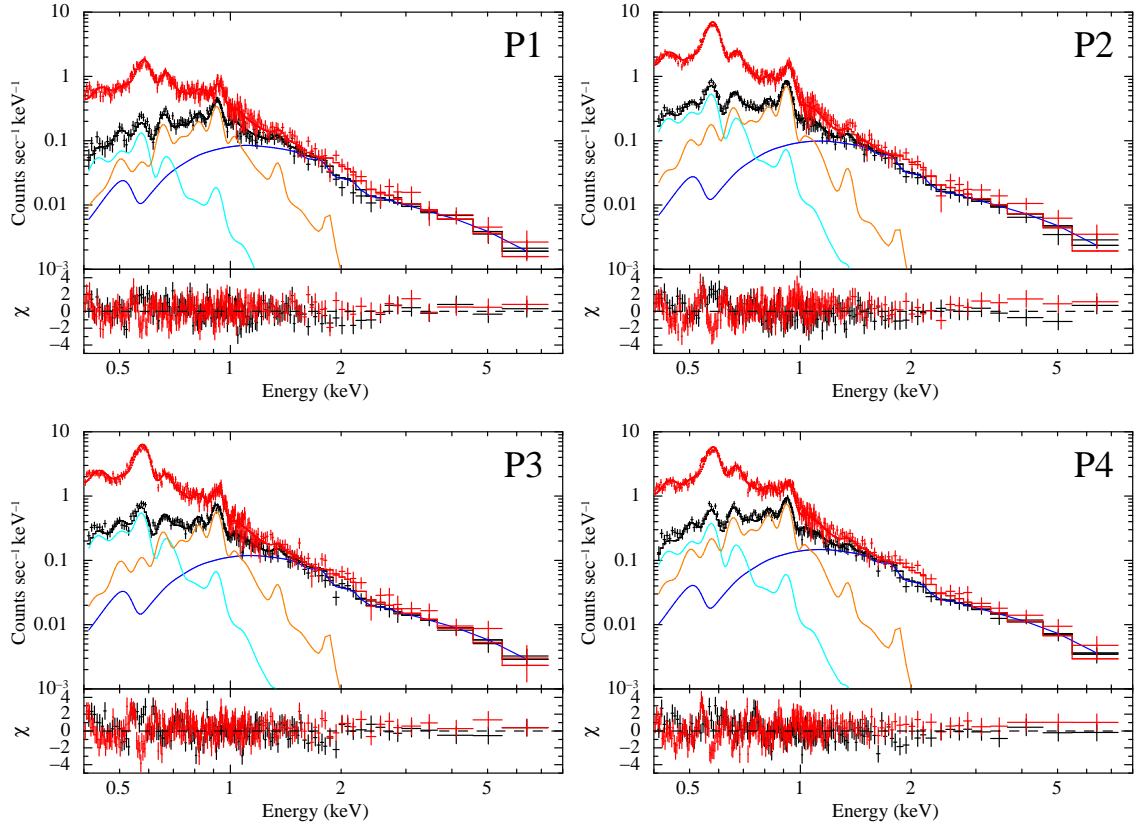}
  \end{center}
  \caption{XIS spectra along with the best-fit model.  Red and
	black correspond to XIS1 and XIS0+3, respectively.
	Contributions of each component ({\tt apec} in 
	light blue, {\tt vnei} in orange, and {\tt power-law} in
	blue) are separately illustrated only for XIS0+3.  Lower 
	panels show residuals between the data and the best-fit model.  
	}
	\label{fig:fit2} 
\end{figure}

\begin{figure}[h]
  \begin{center}
    \FigureFile(150mm,150mm){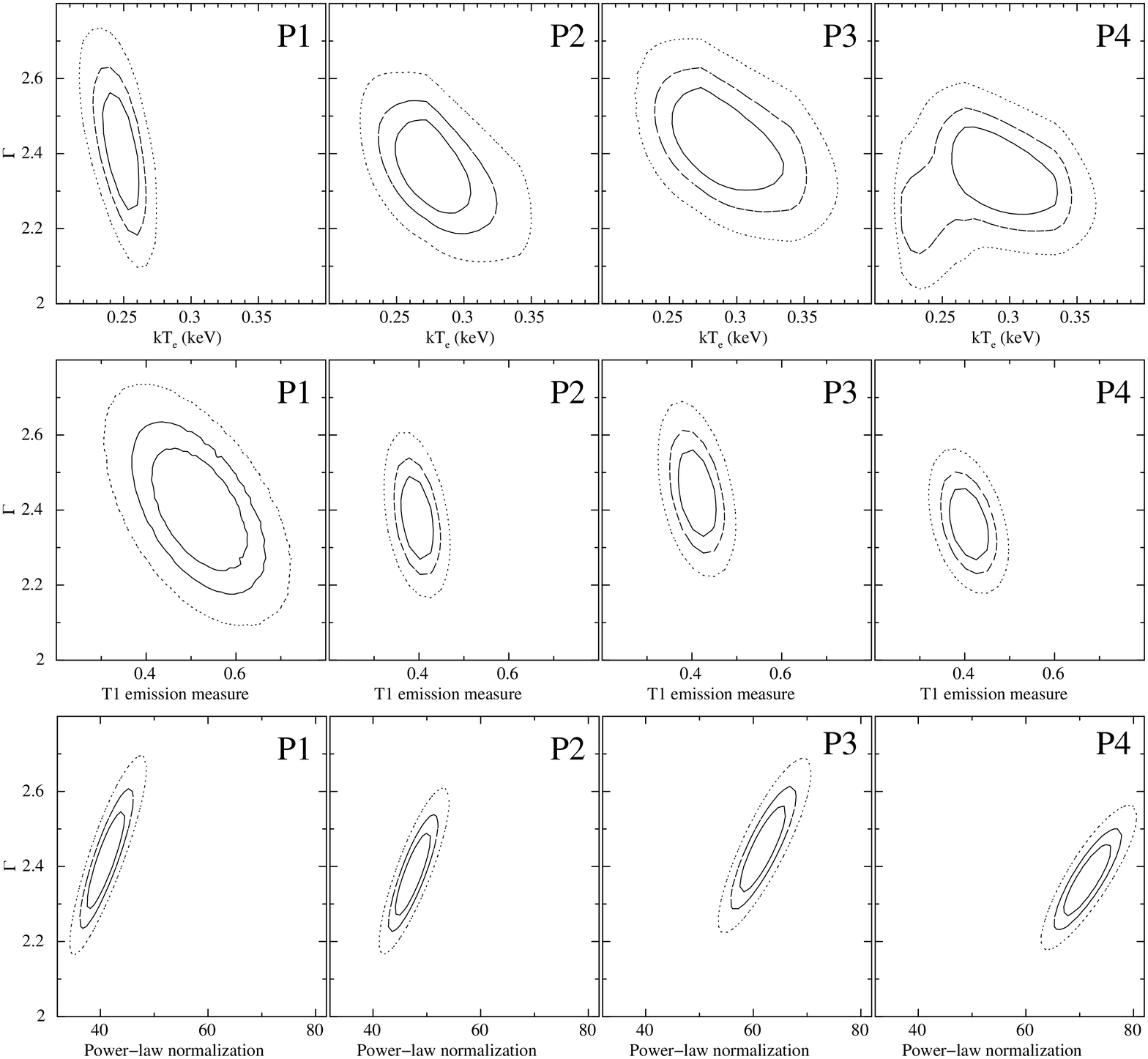}
  \end{center}
  \caption{Confidence contours at 68\% (solid), 90\% (dashed), and
	99\% (dotted) for $\Gamma$ vs.\ the electron temperature in T1 
	(top panels), $\Gamma$ vs.\ the emission measure in T1
	(middle panels), and $\Gamma$ vs.\ normalization in the {\tt
	power-law} component (bottom panels).  We use the {\tt vnei}
	model for the T1 component.  For the middle and
	bottom panels, we calculate the significance level after
	fixing the electron temperature and the ionization timescale
	in T1.  Emission measures in the middle panels are in units of 
	$10^{18}$cm$^{-5}$. Normalizations in the bottom panels are in
	unit of photons\,keV$^{-1}$\,cm$^{-2}$\,s$^{-1}$\,str$^{-1}$
	at 1\,keV.
	}
	\label{fig:conf_cont} 
\end{figure}

\begin{figure}[h]
  \begin{center}
    \FigureFile(85mm,85mm){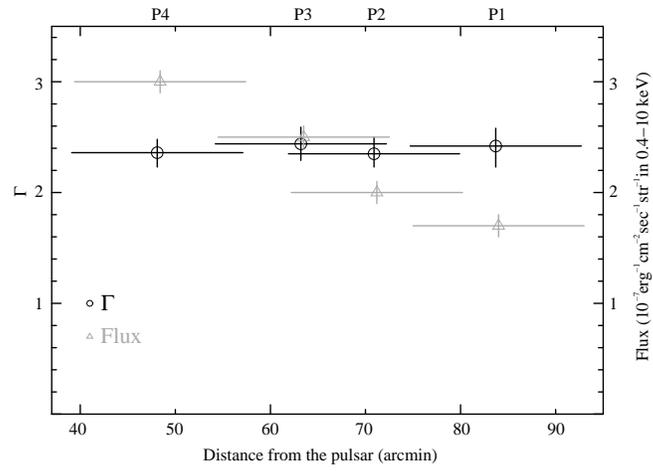}
  \end{center}
  \caption{Photon indices (black circles) and fluxes (grey triangles)
	of the power-law component as a function of the distance from
	the Vela pulsar.  The data points for the photon indices and 
	fluxes are slightly displaces horizontally for clarity.
  }
	\label{fig:index_flux} 
\end{figure}

\begin{table*}[htb]
\begin{center}
\caption{Spectral-fit parameters using a two-thermal-components plus one-powerlaw-component model.$^*$}
\label{tab:fit_param}
\scriptsize\begin{tabular}{lcccccccc}
\hline\hline
Parameter&\multicolumn{2}{c}{P1}&\multicolumn{2}{c}{P2}&\multicolumn{2}{c}{P3}&\multicolumn{2}{c}{P4}\\
\hline
\multicolumn{2}{l}{Interstellar absorption: {\tt tbabs}} &&&&&&&\\
$N_\mathrm{H}$ ($10^{20}$\,cm$^{-2}$)\dotfill & \multicolumn{8}{c}{3 (fixed)}  \\
\hline
\multicolumn{5}{l}{High-temperature component: {\tt vnei} or {\tt vpshock}} \\
& {\tt vnei} & {\tt vpshock}& {\tt vnei} & {\tt vpshock}& {\tt vnei} & {\tt vpshock}& {\tt vnei} & {\tt vpshock}\\
$kT_\mathrm e$ (keV)\dotfill & 0.25$^{+0.02}_{-0.01}$ & 0.24$^{+0.03}_{-0.01}$ & 0.27$^{+0.04}_{-0.02}$ & 0.27$\pm$0.03 & 0.28$^{+0.06}_{-0.03}$ & 0.30$\pm$0.05 & 0.29$^{+0.06}_{-0.03}$ & 0.30$^{+0.05}_{-0.04}$ \\
O (solar)\dotfill & 0.28$^{+0.06}_{-0.06}$ & 0.27$^{+0.02}_{-0.02}$ & 0.52$^{+0.05}_{-0.03}$ & 0.45$^{+0.02}_{-0.05}$ & 0.45$^{+0.06}_{-0.06}$ & 0.39$^{+0.02}_{-0.03}$ & 0.49$^{+0.06}_{-0.07}$ & 0.45$^{+0.04}_{-0.05}$ \\
Ne (solar)\dotfill & 0.62$\pm$0.16 & 0.64$^{+0.12}_{-0.04}$ & 1.14$^{+0.15}_{-0.18}$ & 0.98$\pm$0.11 & 0.81$^{+0.15}_{-0.16}$ & 0.65$^{+0.08}_{-0.05}$ & 0.98$^{+0.15}_{-0.18}$ & 0.80$^{+0.13}_{-0.05}$ \\
Mg (solar)\dotfill & 0.43$^{+0.22}_{-0.2}$ & 0.45$\pm$0.2 & 0.39$\pm$0.2 & 0.36$^{+0.19}_{-0.17}$ & 0.29$^{+0.22}_{-0.21}$ & 0.28$^{+0.18}_{-0.16}$ & 0.29$\pm$0.19 & 0.26$^{+0.16}_{-0.15}$ \\
Fe (solar)\dotfill & 0.29$^{+0.06}_{-0.10}$ & 0.31$^{+0.04}_{-0.03}$ & 0.53$^{+0.09}_{-0.10}$ & 0.53$^{+0.04}_{-0.07}$ & 0.50$^{+0.11}_{-0.12}$ & 0.49$^{+0.05}_{-0.06}$ & 0.59$^{+0.11}_{-0.12}$ & 0.53$^{+0.05}_{-0.06}$ \\
log\{$\tau$(cm$^{-3}$\,sec)\} \dotfill & $>$12  & ---  & 10.89$\pm$0.16  & ---  & 10.78$^{+0.18}_{-0.22}$  & ---  & 11.02$^{+0.36}_{-0.25}$  & --- \\
$\tau_\mathrm{lower}$ (cm$^{-3}$\,sec) \dotfill & --- & 0 (fixed)& --- & 0 (fixed) & --- & 0 (fixed) & --- & 0 (fixed) \\
log\{$\tau_\mathrm{upper}$(cm$^{-3}$\,sec)\} \dotfill & ---  & $>$12  & --- & 11.26$^{+0.18}_{-0.15}$  & --- & 11.00$^{+0.09}_{-0.19}$  & --- & 11.25$^{+0.32}_{-0.22}$ \\
EM$^\dagger$ ($10^{18}$cm$^{-5}$)\dotfill & 0.51$^{+0.12}_{-0.09}$ & 0.52$^{+0.06}_{-0.02}$ & 0.39$^{+0.10}_{-0.09}$ & 0.48$^{+0.03}_{-0.06}$ & 0.41$^{+0.12}_{-0.11}$ & 0.45$^{+0.03}_{-0.09}$ & 0.41$^{+0.17}_{-0.11}$ & 0.46$^{+0.02}_{-0.05}$ \\
Observed flux$^\ddagger$ \dotfill & 2.1$\pm$0.1& 2.1$\pm$0.1& 6.8$\pm$0.3& 7.2$\pm$0.2& 7.0$\pm$0.3& 7.8$\pm$0.2& 6.5$\pm$0.3& 7.9$\pm$0.2 \\
Unabsorbed flux$^\ddagger$ \dotfill & 2.7$\pm$0.1 & 2.6$\pm$0.1 & 8.7$\pm$0.3 & 9$\pm$0.2 & 7.9$\pm$0.3 & 9.8$\pm$0.3 & 6.8$\pm$0.3 & 9.8$\pm$0.2 \\
\hline
\multicolumn{3}{l}{Low-temperature component: {\tt apec}} &&&&&&\\
$kT_\mathrm e$ (keV)\dotfill & 0.09$\pm$0 & 0.09$\pm$0.01 & 0.09$\pm$0.01 & 0.09$\pm$0.01 & 0.08$\pm$0.01 & 0.08$\pm$0.01 & 0.09$\pm$0.01 & 0.09$\pm$0.01 \\
EM$^\dagger$ ($10^{18}$cm$^{-5}$)\dotfill & 1.7$\pm$0.3 & 1.6$^{+0.4}_{-0.2}$ & 7.3$^{+0.8}_{-0.9}$ & 6.2$\pm$0.6 & 13.1$^{+4.8}_{-5.0}$ & 7.4$^{+5.1}_{-1.8}$ & 4.7$^{+0.6}_{-0.7}$ & 3.4$^{+0.7}_{-0.8}$ \\
Observed flux$^\ddagger$ \dotfill & 2.1$\pm$0.1& 2.0$\pm$0.1& 6.6$\pm$0.3& 6.3$\pm$0.3& 5.9$\pm$0.3& 5.3$\pm$0.2& 5.2$\pm$0.3& 3.9$\pm$0.3 \\
Unabsorbed flux$^\ddagger$ \dotfill & 2.7$\pm$0.1 & 2.6$\pm$0.1 & 8.7$\pm$0.3 & 8.4$\pm$0.3 & 7.9$\pm$0.3 & 7$\pm$0.3 & 6.8$\pm$0.3 & 5.2$\pm$0.3 \\
\hline
\multicolumn{3}{l}{Hard-tail component: {\tt power-law}} &&&&&&\\
$\Gamma$ \dotfill & 2.42$^{+0.16}_{-0.19}$ & 2.34$^{+0.17}_{-0.06}$ & 2.35$^{+0.15}_{-0.12}$ & 2.33$^{+0.12}_{-0.11}$ & 2.44$\pm$0.15 & 2.38$^{+0.10}_{-0.17}$ & 2.36$^{+0.12}_{-0.13}$ & 2.28$^{+0.14}_{-0.08}$ \\
Norm$^\S$ \dotfill & 41.0$^{+4.8}_{-3.0}$ & 38.3$^{+1.8}_{-3.1}$ & 46.6$^{+4.3}_{-4.5}$ & 45.5$^{+3.1}_{-2.5}$ & 61.6$^{+5.5}_{-3.4}$ & 57.3$^{+2.7}_{-6.6}$ & 71.2$^{+6.2}_{-7.2}$ & 67.9$^{+3.6}_{-3.1}$ \\
Observed flux$^\ddagger$ \dotfill & 1.6$\pm$0.1& 1.6$\pm$0.1& 1.8$\pm$0.1& 1.8$\pm$0.1& 2.3$\pm$0.1& 2.2$\pm$0.1& 2.8$\pm$0.1& 2.7$\pm$0.1 \\
Unabsorbed flux$^\ddagger$ \dotfill & 1.7$\pm$0.1 & 1.7$\pm$0.1 & 2$\pm$0.1 & 2$\pm$0.1 & 2.5$\pm$0.1 & 2.4$\pm$0.1 & 3$\pm$0.1 & 2.9$\pm$0.1 \\
\hline
Offset (eV)\dotfill & +0.9, +10.9 & +0.9, +10.8 & 0.0, +6.9 & 0.0, +7.2 & 0.0, +6.9 & +0.2, +6.0 & +0.4, +8.5 & +0.3, +8.0 \\
\hline
$\chi^{2}$/d.o.f. \dotfill & 400/387 & 401/387 & 511/387 & 520/387 & 538/387 & 536/387 & 507/387 & 506/387 \\
\hline
\\[-5pt]
\multicolumn{9}{@{}l@{}}{\hbox to 0pt{\parbox{168mm}{\footnotesize
\par\noindent
\footnotemark[$^*$]Other elements are fixed to the solar values. The values of abundances are multiples of solar values.  The errors are in the range $\Delta\,\chi^2\,<\,2.7$.  The fixed $N_\mathrm{H}$-value is determined by fitting the Vela pulsar's spectrum with a blackbody model (Mori et al. in preparation).
\par\noindent
\footnotemark[$^\dagger$]EM denotes the emission measure $\int n_\mathrm{e} n_\mathrm{H} d\ell$, where $n_\mathrm{H}$ is the number density of protons and $d\ell$ is the X-ray--emitting plasma depth.
\par\noindent
\footnotemark[$^\ddagger$]In unit of 10$^{-7}$ erg\,cm$^{-2}$\,sec$^{-1}$\,str$^{-1}$ in a range of 0.4--10\,keV.
\par\noindent
\footnotemark[$^\S$]In unit of photons\,keV$^{-1}$\,cm$^{-2}$\,s$^{-1}$\,str$^{-1}$ at 1\,keV.
}\hss}}

\end{tabular}
\end{center}
\end{table*}
\end{document}